\documentclass[conference]{IEEEtran}
\IEEEoverridecommandlockouts
\usepackage{cite}
\usepackage{amsmath,amssymb,amsfonts}
\usepackage{algorithmic}
\usepackage{graphicx}
\usepackage{textcomp}
\usepackage{caption}
\usepackage{subcaption}
\usepackage[svgnames]{xcolor}
\def\BibTeX{{\rm B\kern-.05em{\sc i\kern-.025em b}\kern-.08em
    T\kern-.1667em\lower.7ex\hbox{E}\kern-.125emX}}

\begin{document}

\title{\title{\huge Graph Analysis of Citation and Co-authorship Networks of Egyptian Authors}
{\footnotesize \textsuperscript{*}Note: Sub-titles are not captured in Xplore and
should not be used}
\thanks{Identify applicable funding agency here. If none, delete this.}
}

\author{\IEEEauthorblockN{Mariam Ayman$^*$, Sohaila Kandil$^*$, Alaa Moheb$^*$, Ahmed Abdelkader$^*$, Mohanned Ahmed$^*$, Walid Gomaa$^{*a}$}
\IEEEauthorblockA{$^*$Department of Computer Engineering, Egypt-Japan University of Science and Technology, Alexandria, Egypt.}
\IEEEauthorblockA{$^a$Faculty of Engineering, Alexandria University, Alexandria, Egypt}
}
\maketitle

\begin{abstract}
The current research conducts a comprehensive analysis of citation networks focusing on publications by authors affiliated with Egyptian institutions. Leveraging the Semantic Scholar platform and its API, a citation network and 
a co-authorship network graphs are constructed to visualize the interconnections among these publications and their authors. This is done using the Python package for graph analysis (Networkx). 
The primary objective is to identify influential Egyptian publications and assess the centrality of nodes within the citation network. Through meticulous data collection including web scraping techniques, we obtained a cleaned dataset comprising publications by authors affiliated with Egyptian institutions. 
The analysis addresses challenges related to data quality, technical intricacies, and time constraints, resulting in a reliable and robust dataset. The findings provide valuable information on the impact of Egyptian publications, offering insights into the scholarly influence of authors associated with Egyptian institutions. This research equips researchers and academics interested in evaluating the impact of Egyptian publications with valuable data for future studies, collaborations, and policy decisions.
\end{abstract}


\section{Introduction}
\label{introduction}

\par 
Citation networks are essential tools for evaluating the influence and impact of scholarly publications across specific research domains \cite{book}. These networks visually represent relationships between publications, where nodes symbolize papers, and edges indicate citation links. Metrics such as citation counts, eigenvector scores, and network centrality measures are commonly used to assess the significance of individual works and authors. Citation networks also help identify research trends, interdisciplinary contributions, and collaborative patterns, providing a comprehensive view of academic influence \cite{DetectingTrends} \cite{Portenoy2017}. Our study focuses on conducting a comprehensive analysis of citation networks for authors affiliated with Egyptian institutions. The primary objective is to identify 
influential publications and authors and evaluate the centrality of nodes within these networks.

\par 
Egyptian publications have made significant contributions to advancing knowledge across various fields, making it essential to understand their impact. By analyzing the citation networks of Egyptian authors, we aim to uncover influential publications and explore the connections that exist among them. This analysis will enable us to identify articles that have garnered a substantial number of citations, indicating their widespread influence within the scholarly community.

\par 
To achieve our goals, we utilized the Semantic Scholar platform and its API, along with Google Scholar's API, to gather data on publications authored by individuals affiliated with Egyptian institutions. Then, we constructed a citation network graph to represent the interconnections among these publications. In this network, each article or paper is represented as a node, and the citation relationships are depicted as edges.

\par 
The findings of this research provide valuable insights into the scholarly influence of authors associated with Egyptian institutions. By identifying influential publications and assessing the centrality of nodes within citation networks, we gain a deeper understanding of their impact. Additionally, the analysis sheds light on authorship patterns and co-authorship relationships, particularly among
Egyptian scholars. Moreover, it reveals certain behaviors adopted by authors to enhance their scholarly reputation.

\par 
The paper is organized as follows. Section II discusses citation and co-authorship networks, while Section III presents a review of related literature. Section IV describes the methodology, including data collection and cleaning. Section V analyzes the citation network with a focus on temporal trends, degree distribution, authorship patterns, and network metrics. Section VI examines co-authorship networks, including centrality and component analysis. Finally, the conclusion and acknowledgments are provided in Sections VII and VIII, respectively.

\section{Citation Network}
\label{Citation network}

\par 
A citation network is a valuable tool for studying the relationships and interconnections among academic papers \cite{martinez2020current}. In this network, each paper or article is represented as a node, and a directed edge exists between two nodes if one paper cites the other. This allows to analyze and visualize the flow of information and references within the scholarly community.
By analyzing the citation network, we can gain insights into the impact of specific papers as well as the influence of certain authors \cite{CitationNetworkAnalysis}. Nodes that receive a significant number of incoming citations are considered influential, as they indicate the widespread recognition and relevance of the research. These influential nodes play a crucial role in shaping the academic discourse and guiding future studies \cite{DetectingTrends}. Citation networks can also reveal patterns of academic collaboration. By examining co-authorship networks, we can identify key researchers who frequently collaborate, forming tight-knit academic communities. This analysis can uncover trends in research partnerships, the flow of ideas across institutions, and potential areas for future interdisciplinary work. Furthermore, analyzing citation frequency and network centrality can help recognize emerging scholars and the evolving focus of research fields \cite{info15100597} \cite{10184723}.

\par
\begin{figure}
	\centering 
	\includegraphics[width=0.4\textwidth]{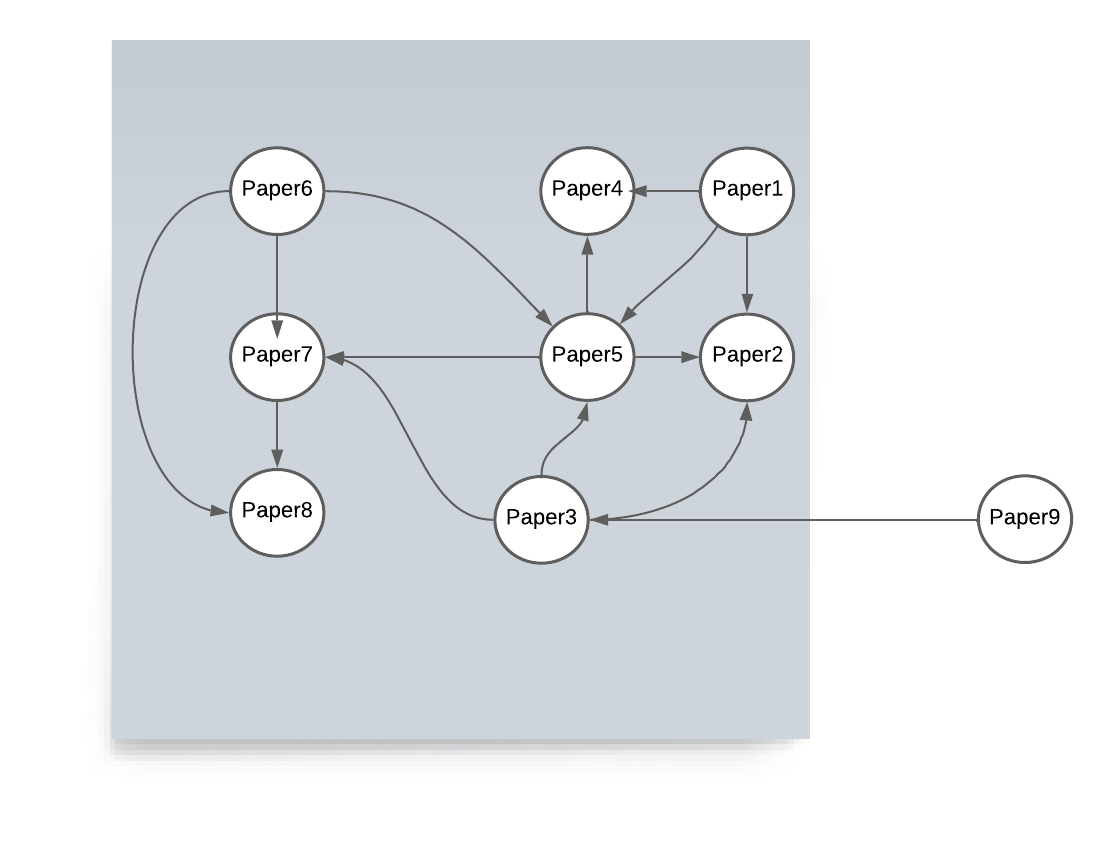}	
	\caption{This image represents a sample citation network. A directed line from Paper A to Paper B indicates that Paper A cites Paper B. Notably, the grey area exhibits relatively high centrality, as each node (representing a paper) is connected to multiple surrounding nodes. In contrast, Paper 9 is located in a region of low connectivity, with minimal connections, making it an uncommon or isolated} 
	\label{fig1}%
\end{figure}

 
In Figure \ref{fig1}, illustrates a sample visualization of a citation network. Each node represents a paper, and the directed edges depict the citation relationships between them. If there is an edge from paper A to paper B, it indicates that paper A cites paper B. This network shows how papers are interconnected through citations, forming a web of knowledge. 

\section{Co-authorship Network}
\label{sec: Co-authorship Network}

\par 
A co-authorship network is a type of social network that focuses on the collaboration relationships between authors. In this network, each author is represented as a node, and an edge connects two authors if they have collaborated on one or more publications together.
The co-authorship network enables the understanding of the patterns of collaboration within the scholarly community \cite{fagan2018assessing}. By examining the connections between authors, we can identify clusters or communities of researchers who frequently work together. These collaborations often indicate shared research interests, expertise, or institutional affiliations \cite{ARIELXU2020100307}.

\par
\begin{figure}
	\centering 
	\includegraphics[width=0.4\textwidth]{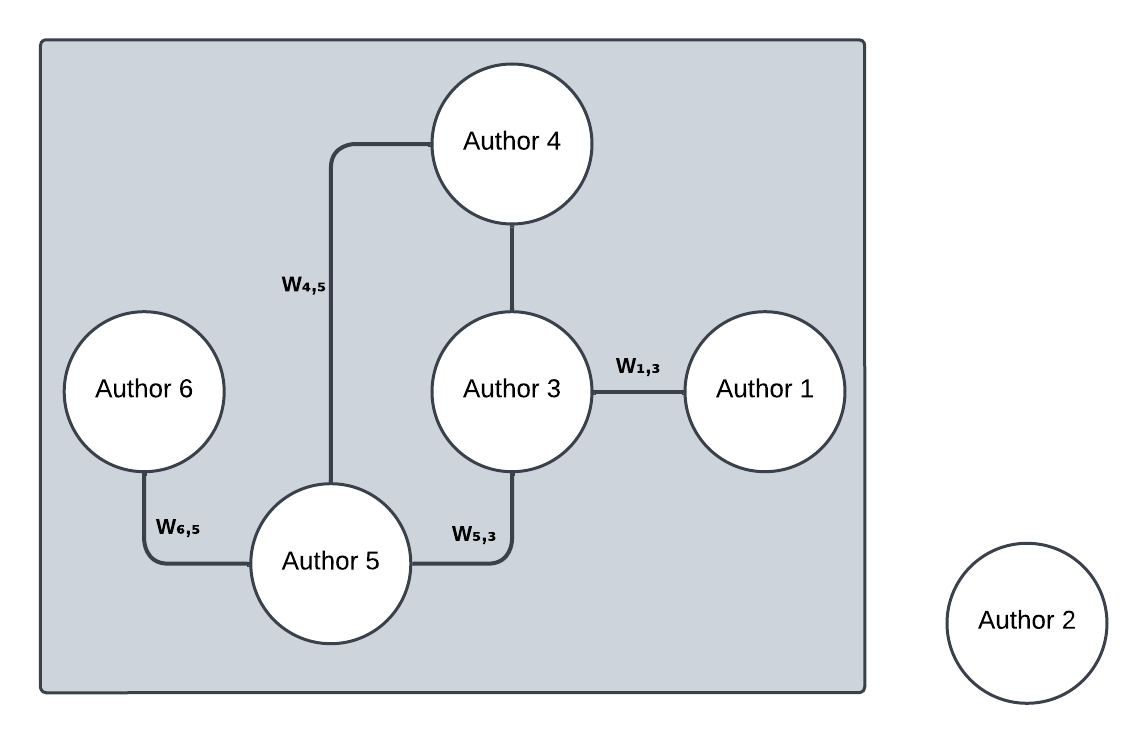}	
	\caption{This diagram represents a co-authorship network, where each node corresponds to an author and a line (edge) between two nodes indicates collaboration between the authors. The weight \( W_{A,B} \) on an edge represents the strength of collaboration between Author \( A \) and Author \( B \). Authors in the grey area exhibit high centrality, indicating that they are well-connected and collaborate with many others. In contrast, Author 2 is located in a low-centrality area, reflecting fewer collaborations within the network.} 
	\label{fig2}%
\end{figure}

 
In Figure \ref{fig2}, we present a visualization of a co-authorship network. Each node represents an author, and the edges represent co-authorship relationships 
between authors who have collaborated on publications. The weight of an edge can be interpreted as the number of times two authors have coauthored publications together, providing a measure of the strength of their
collaboration. This network is naturally undirected as co-authorship is a symmetric relation, unlike the citation network which has an inherent order induced by citation direction.

\par 
Studying the co-authorship network helps us identify influential researchers and research groups who have established extensive collaborative networks. It also provides insights into the dynamics of knowledge production, as authors with strong collaborative ties may have a higher likelihood of producing impactful research.

\par 
Both the citation network and co-authorship network offer valuable perspectives on the scholarly landscape associated with Egyptian institutions. By analyzing these networks and assessing various network metrics using tools like Networkx, we can gain a deeper understanding of the impact, influence, and collaboration patterns within the Egyptian academic community. These insights are essential for researchers, academics, and policymakers interested in evaluating the scholarly contributions of Egyptian publications and fostering future collaborations and policy decisions. It is a case study that can be extended widely.

\section{Literature Review}

\par
V. Umadevi~\cite{umadevi2013case} explores the Co-authorship networks of published papers, which depict collaborations among researchers involved in scholarly publications. In this context, the Co-authorship network is established by representing researchers as nodes, while connections between nodes (edges) symbolize collaborative efforts between researchers. To facilitate the visualization and analysis of these networks, the study employs the Gephi tool, a widely utilized software specifically designed for network visualization and exploration.

\par 
Azimjonov et. al.~\cite{DBLP:journals/corr/abs-1807-09009} present a methodology for collecting metadata from papers, including titles, abstracts, keywords, body text, conclusion, and references. They utilized targeted keyword searches and manual extraction processes to extract the required metadata. To enhance the extraction process, they utilized Java programming language and Optical Character Recognition (OCR) libraries.

\par 
Our study improves upon this approach by employing web scraping techniques on the Google Scholar website to collect data on Egyptian authors. Additionally, we utilize the Semantic Scholar API to retrieve a comprehensive dataset of papers associated with each researcher identified through web scraping.

\section{Methodology}
\subsection{Data Collection}

\par
The dataset focuses on Egyptian-authored papers, defined as publications where at least one author is either Egyptian or affiliated with an Egyptian university. For each paper, the dataset includes the title, a unique paper ID, the publication date, a list of authors along with their names and IDs, the number of citations, and a list of references with their titles and IDs.
To build our dataset, we followed a two-stage process. In the first stage, we identified Egyptian researchers by collecting information from Google Scholar, determining an author’s affiliation with an Egyptian university as the criterion for inclusion. In the second stage, we utilized the Semantic Scholar API to gather data on papers authored by these researchers. Semantic Scholar, an AI-powered research platform, provided comprehensive access to academic publications. The specific details of each stage are discussed in the following sections.

\par
In phase 1, we utilized web scraping techniques on the Google Scholar website to gather data on researchers affiliated with prominent Egyptian universities. The universities included in this process were Ain Shams University, Cairo University, Egypt-Japan University of Science and Technology (E-JUST), Alexandria University, Banha University, Assiut University, and Zewail University. This effort resulted in a dataset comprising 13,027 entries, which included researchers' names and their affiliations. 
The scraping process was implemented using the Selenium library in Python. To adhere to Google Scholar’s guidelines and avoid exceeding permissible data scraping rates, we adopted a structured approach. Data for each university was collected in separate sessions, with sufficient time intervals between scraping instances to ensure compliance and minimize the risk of interruptions.

\par
In phase 2, we utilized the Semantic Scholar database to gather data of research papers each published by each of the  researchers identified in phase 1. 
Using API requests with a private Semantic Scholar API key, we followed the documented guidelines provided by Semantic Scholar \cite{semantic-scholar-docs}. For each researcher, we sent an API request by name to retrieve data on their published papers, including titles, paper IDs, and publication dates. Subsequently, we made additional API requests for each paper using its Semantic Scholar ID to obtain detailed information, such as references, the number of citations, and the names and Semantic Scholar IDs of all co-authors associated with the publication.

\par
The final dataset, collected in phase 2, is stored in a CSV file and consists of 31,508 research papers with a total of 320,969 citations. This dataset, named "AlGoNet", will serve as the foundation for future analyses and studies.

\subsection{Data Cleaning}

\par
Data cleaning was carried out in two stages to ensure the dataset's accuracy and consistency. In the first stage, a validation step was incorporated during the data collection process. This step utilized a code segment to identify and exclude duplicate entries before adding them to the CSV file.
In the second stage, we used the Pandas library in Python to further refine the dataset. This involved systematically removing any remaining duplicates and empty fields. These measures ensured the dataset's integrity, preparing it for subsequent analysis with minimal noise or inconsistencies.

\subsection{Analysis of Citation Network}

\subsubsection{Temporal Analysis of Publications}
\par
\begin{figure}
	\centering 
	\includegraphics[width=0.4\textwidth]{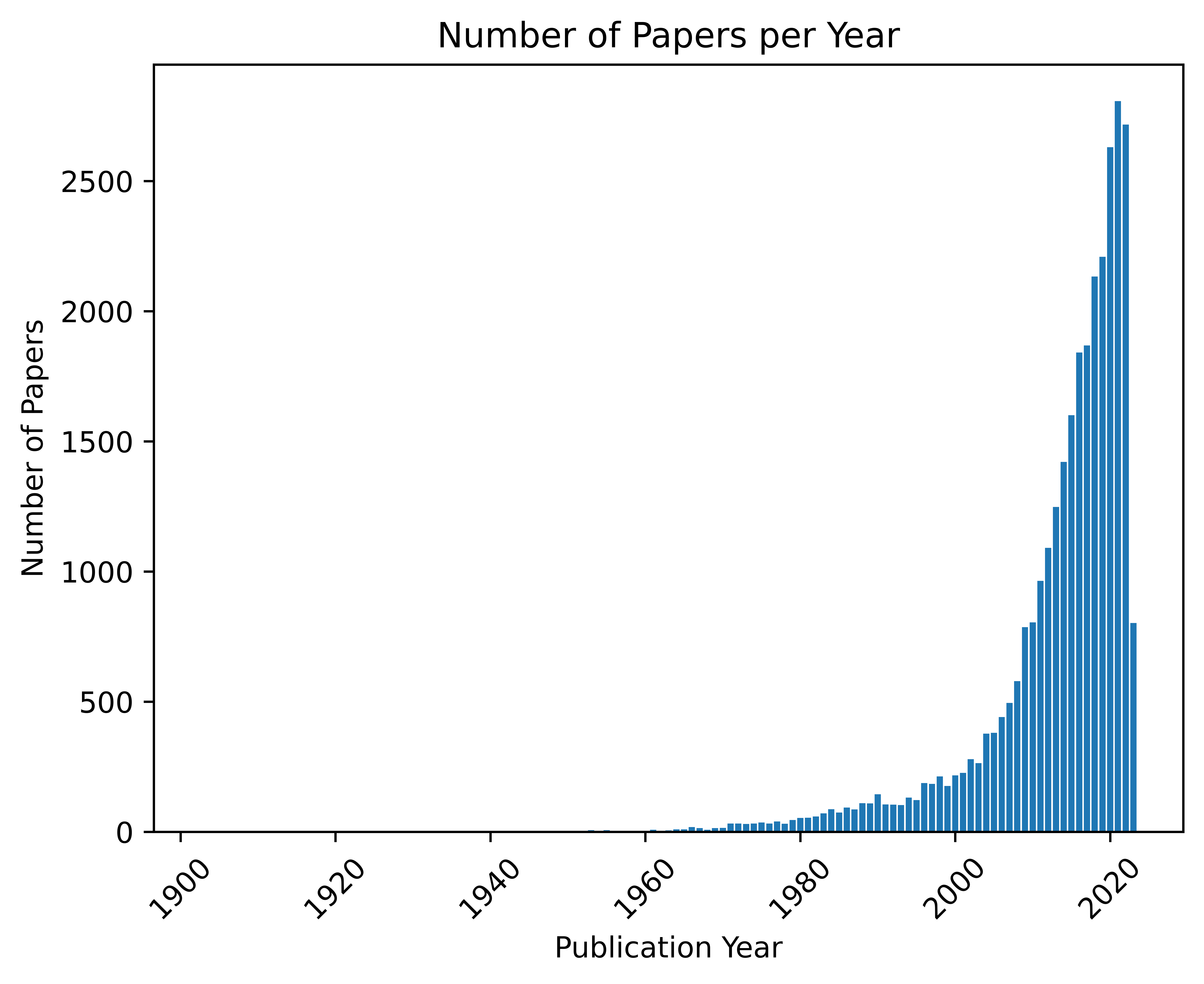}	
	\caption{Publications distribution over the years. } 
	\label{temporal}%
\end{figure}

 
Analyzing our dataset to get insights about the publication trend, we found that the oldest paper was published in 1903 while the most recent is in 2023.
As shown from the trend in Figure \ref{temporal}, most of the manuscripts were recently published within the years from 2000 to 2023.

\subsubsection{Degree Distribution Analysis}

\par
\begin{figure}
	\centering 
	\includegraphics[width=0.4\textwidth]{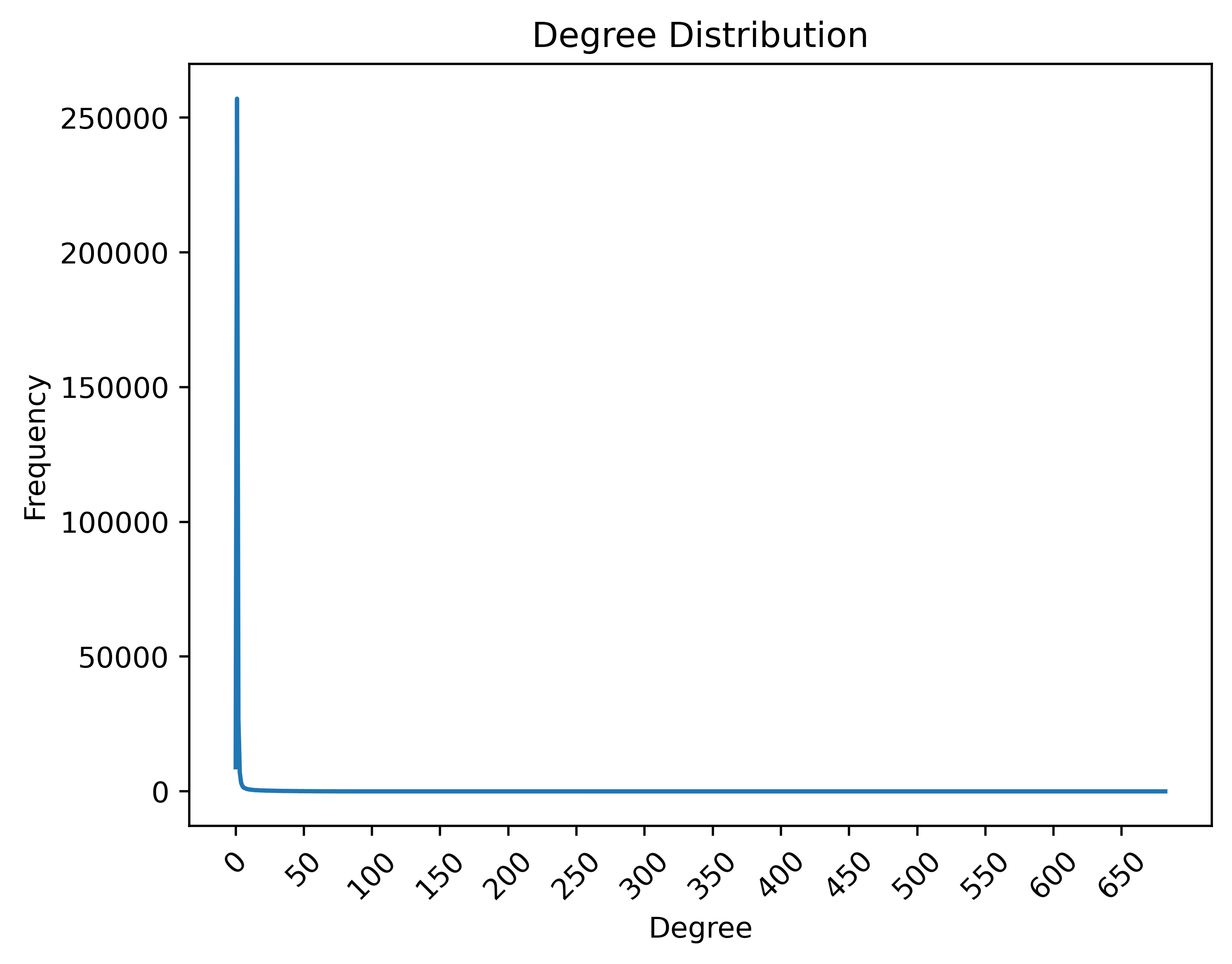}	
	\caption{Node degree distribution. } 
	\label{distribution}%
\end{figure}

\begin{figure}[!ht]
     \centering
     \begin{subfigure}[b]{0.4\textwidth}
         \centering
         \includegraphics[width=\textwidth]{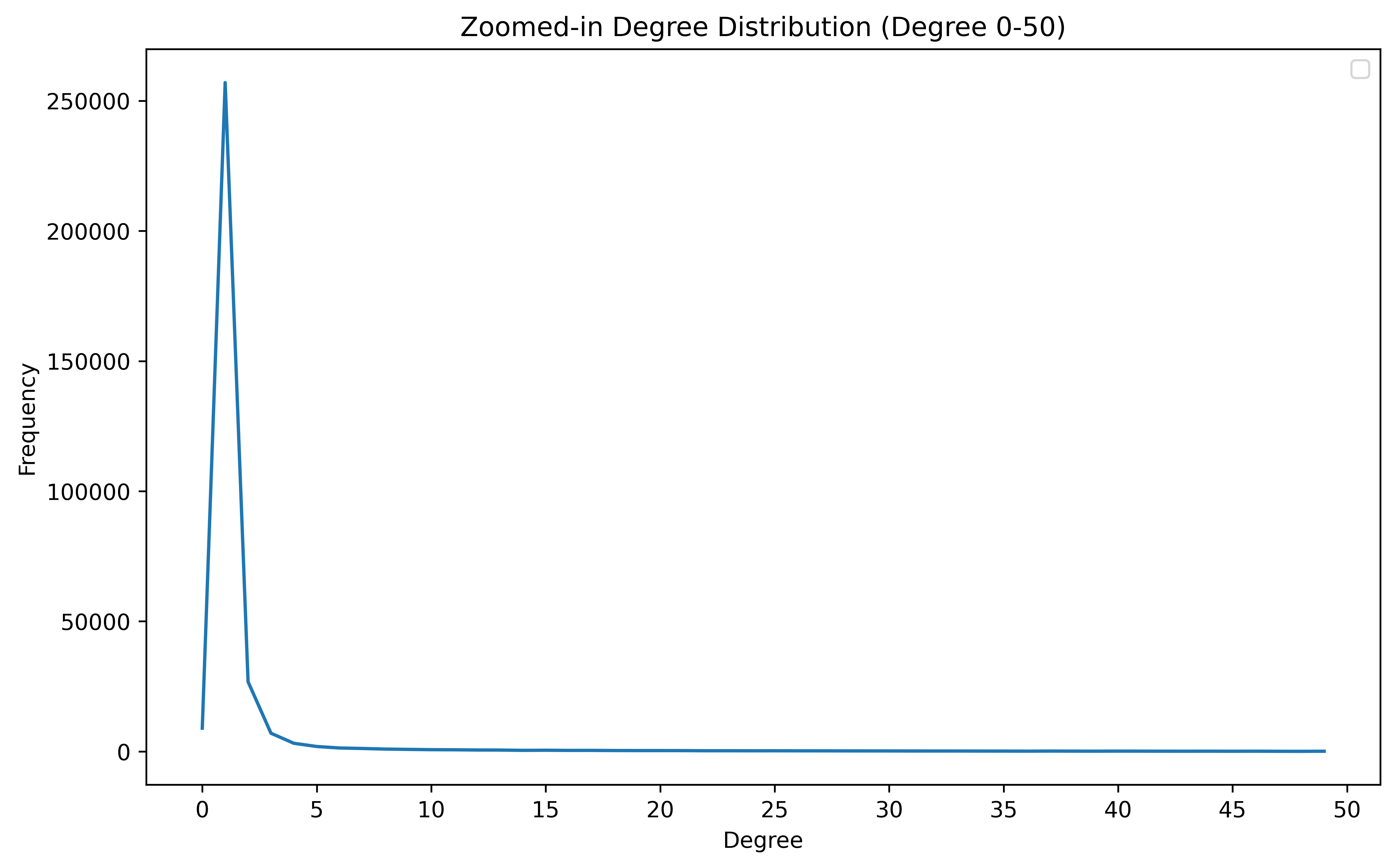}
          \caption{Degree range 0-50}
         \label{fig:part_a}
     \end{subfigure}
     \hfill
     \begin{subfigure}[b]{0.4\textwidth}
         \centering
         \includegraphics[width=\textwidth]{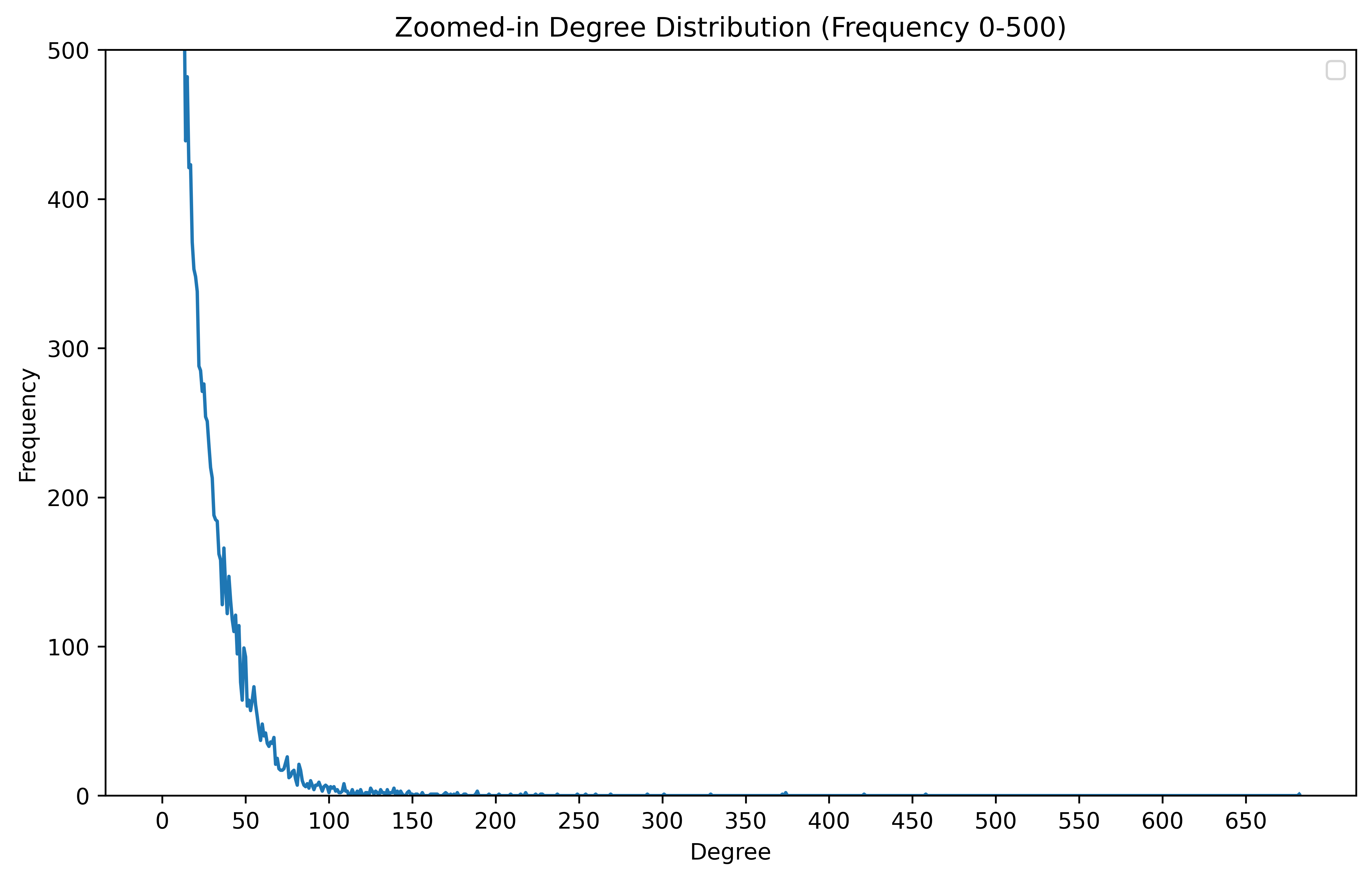}
         \caption{Frequency range 0-500}
         \label{fig:part_b}
     \end{subfigure}
        \caption{Zoomed-in degree distributions: (a) Degree range 0-50, (b) Frequency range 0-500.}
        \label{fig:distribution_zoom}
\end{figure}

 
The degree distribution shows how the number of nodes in the network is distributed across different degrees where the degree of a node represents the number of citations it received from other nodes in the network. 
As Figure \ref{distribution} shows, the network has a sharp peak at lower degrees, indicating that most nodes (publications) have a low number of connections (citations), while a small number of nodes are highly influential.
Figure \ref{fig:part_a} shows a zooming-in view of the degree distribution graph for the range 0-50. From the graph, it can be concluded that the majority of nodes in the network have a very low degree (less than 5) with the highest peak at the degree of 1. This indicates that most publications by Egyptian authors have a relatively low number of citations, by other Egyptian researchers.
Figure \ref{fig:part_b} represents a zoomed-in degree distribution of the frequency range from 0-500, which is the small number of highly cited publications that have a much higher degree. As observed, the frequency of the degrees after 150 starts to get quite low, indicating that highly cited publications are rare.
The degree distribution also appears to follow a power law distribution, with a few highly connected nodes and many nodes with few connections. The power-law distribution can be expressed as \( P(x) = Cx^{-\alpha} \), where \( \alpha \) is the scaling exponent, (\( x_{\text{min}} \)) is the best minimal value for power law fit and \( C \) is a normalization constant. Using the Python \texttt{powerlaw} library, the scaling exponent \( \alpha \) was estimated to be \( 1.7259583924156112 \), with a minimum degree threshold (\( x_{\text{min}} \)) of \( 158 \).
This pattern is often observed in citation networks \cite{Redner_1998} and is known as the ``rich get richer'' phenomenon, where highly cited publications tend to receive more citations over time, leading to a skewed distribution of 
citations. Overall, the degree distribution analysis suggests that there are few highly influential publications, by Egyptian authors, that have received a large number of citations, while the majority of publications, surely by Egyptian authors too, have received relatively few citations.

\subsubsection{Authorship Pattern Analysis}

\par
\begin{figure}
	\centering 
	\includegraphics[width=0.4\textwidth]{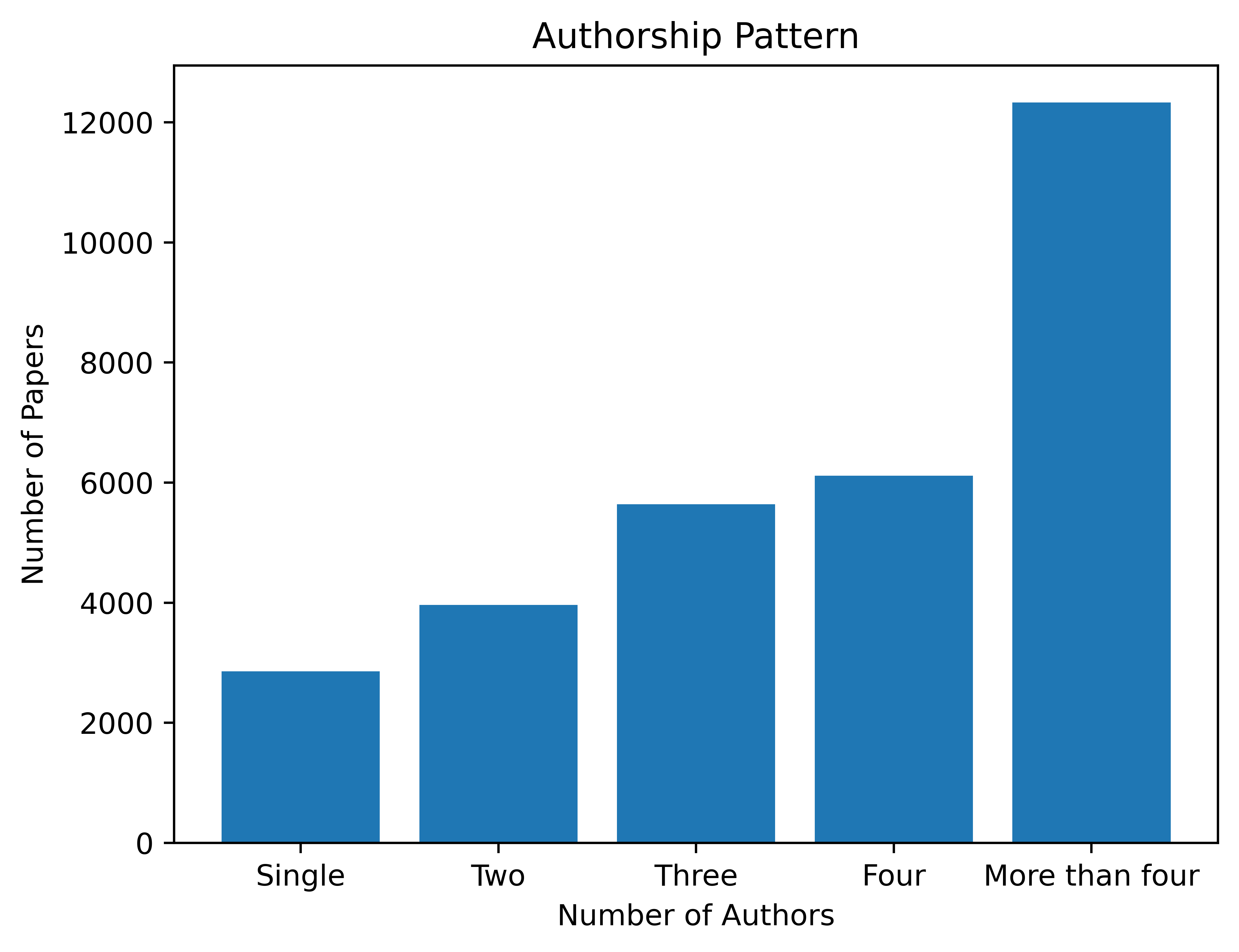}	
	\caption{Authorship pattern analysis.} 
	\label{fig5}%
\end{figure}

 
It is evident from Figure \ref{fig5} that out of all the included papers (30,905 papers), papers with four authors (6,115 papers) are little ahead than three-authored 
papers (5,640 papers) followed by two-authored papers (3,964 papers), while single-authored papers (2,856 articles) are at the back foot. The remaining papers having more than four authors (12,330 papers) are the 
majority keeping in mind that the limit of the API used in collecting the dataset was 500 authors per paper. 
Therefore, although the maximum author count is 500 authors for a total of 46 papers, those papers may include more than 500 authors.
Hence, it is inferred that, the trend of collaborative research has taken place among the Egyptian papers of this collected dataset.

\subsubsection{Clustering Coefficients}
The \textit{clustering coefficient} is a measure of the tendency of nodes in a network to form tightly connected groups or "triangles" \cite{article}. For an individual node, the clustering coefficient is calculated as the ratio of the number of edges between its neighbors to the total possible number of such edges. Mathematically, for a node \(i\), the local clustering coefficient \(C_i\) is given by:
\[
C_i = \frac{2 \times e_i}{k_i \times (k_i - 1)}
\]
where \(e_i\) is the number of edges between the \(k_i\) neighbors of node \(i\), and \(k_i\) represents the degree of the node. The average clustering coefficient of the network is the mean of the local clustering coefficients of all nodes.
In the current context, the citation network under analysis exhibits a low average clustering coefficient (\(C = 0.019\)). This suggests that the nodes (academic papers) in the network have minimal local interconnectedness, indicating sparse connections among the neighbors of each node. The absence of tightly connected communities or clusters implies that:

\begin{itemize}
    \item \textbf{Diverse Topics:} The papers in the citation network may span multiple research areas, resulting in limited overlap in citations.
    \item \textbf{Sparse Citations:} Citations between papers may occur in a more random or sporadic fashion, without forming cohesive groups of interrelated research.
    \item \textbf{Cross-disciplinary Nature:} The network might include papers from various disciplines, which weakly connect instead of forming dense clusters.
\end{itemize}

\subsubsection{Density Analysis}

The \textit{density} of a graph provides a measure of how many edges exist in the network relative to the maximum possible number of edges. Intuitively, it quantifies the overall "connectedness" of the network. Mathematically, the density \(D\) of a graph \(G = (V, E)\) is defined as:

\[
D = \frac{2 \times |E|}{|V| \times (|V| - 1)}
\]

where \(|V|\) represents the number of nodes, and \(|E|\) is the number of edges in the graph. The formula accounts for all possible undirected connections between nodes in the network.

In our analysis, the density of the citation network is very low (\(D = 3.55 \times 10^{-6}\)), which indicates that the network is sparsely connected. This means the actual number of citation links is negligible compared to the total potential connections. The low density suggests several possible explanations:
\begin{itemize}
    \item Many papers in the network may have few or no citations.
    \item The network likely spans a wide range of topics, reducing the likelihood of dense interconnections between nodes.
    \item Larger networks, such as this one, inherently tend to exhibit lower density because the number of possible connections increases quadratically with the number of nodes, while the actual edges grow at a slower rate.
\end{itemize}

\subsubsection{Centrality Measures}
\textit{Centrality measures} are used to identify the most important nodes in a network based on specific criteria. Intuitively, these measures evaluate the role of a node in the structure and function of the network. Two commonly used centrality measures are:

\begin{itemize}
    \item \textbf{Degree Centrality:} This measures the number of direct connections (edges) a node has. Mathematically, the degree centrality \(C_D(i)\) for a node \(i\) is given by:
    \[
    C_D(i) = \frac{\text{deg}(i)}{|V| - 1}
    \]
    where \(\text{deg}(i)\) is the degree of the node (number of edges connected to \(i\)).

    \item \textbf{Eigenvector Centrality:} This measures a node's influence based on both the quantity and quality of its connections. Nodes connected to highly influential nodes receive higher eigenvector centrality scores. The eigenvector centrality \(C_E(i)\) is derived from the eigenvector equation:
    \[
    \mathbf{C_E} = \lambda \mathbf{A} \mathbf{C_E}
    \]
    where \(\mathbf{A}\) is the adjacency matrix of the graph, \(\lambda\) is the eigenvalue, and \(\mathbf{C_E}\) is the eigenvector of \(\mathbf{A}\).

\end{itemize}

\par 
Computing the degree centrality measure for all the included publications, we got an average degree of 2.8 with the highest degree (= 682) for the article (volumes 357-377 )(1995) \cite{ahmad1995}. 
Interestingly, approximately two-thirds of the publications have a degree of zero or very close to zero, meaning these papers have minimal or no direct connections to others. This suggests a lack of visibility or integration of these works within the scholarly network.

For further analysis, we computed degree and eigenvector centrality measures for recent publications published after 2015, a total of 17007 papers.

\paragraph{\textbf{Degree Centrality for Publications (from 2016 onward)}}
Degree centrality measures the number of direct connections a node has, i.e., the number of citations a paper has. It provides a measure of how influential a paper is based on the number of direct references it receives. Of all the 17007 papers, being published from 2016 onward, the highest degree obtained (=0.0015) was for the publication: ``Managing gsh elevation and hypoxia to overcome resistance of cancer therapies using functionalized nanocarriers, published in 2021'' \cite{DABBOUR2022103022}. This highlights the paper's relatively high level of direct influence within the examined network of recent publications.

\paragraph{\textbf{Eigenvector Centrality for Publications (from 2016 onward)}} 
Eigenvector centrality extends beyond local connections by considering both the number of citations a paper receives and the influence of the citing papers. Unlike degree centrality, which measures direct citations (local influence), eigenvector centrality assigns greater importance to papers cited by other highly central (influential) works, providing a measure of global connectivity and recognition.

The publication “Efficacy and Biological Correlates of Response in a Phase II Study of Venetoclax Monotherapy in Patients with Acute Myelogenous Leukemia,” published in 2016 \cite{konopleva2016efficacy}, achieved the highest eigenvector centrality score (0.314) while also attaining a degree centrality of 510 (citations). In comparison, the paper with the highest degree centrality, "Volumes 357–377 (1995)" \cite{ahmad1995}, achieved a degree centrality of 682 but did not exhibit a comparable eigenvector centrality score. This difference indicates that while the earlier paper is highly cited, its citations are less connected to other influential papers, thus limiting its global influence within the network. In contrast, the Venetoclax study not only garnered significant direct citations but was also referenced by other highly influential works, amplifying its importance across the network.

The prominence of the Venetoclax study can be attributed to its groundbreaking contributions to cancer research, particularly in the treatment of Acute Myelogenous Leukemia using Venetoclax monotherapy. The paper’s clinical relevance and impact on subsequent oncology studies have made it a cornerstone publication. Its high recognition is further supported by the affiliations of its authors with leading institutions, such as The University of Texas MD Anderson Cancer Center, Dana-Farber Cancer Institute, and The Ohio State University Comprehensive Cancer Center, alongside industry collaborators from AbbVie, Inc. These factors collectively explain the study’s high eigenvector centrality, reflecting both its local influence and its strong global connectivity within the citation network.

\subsubsection{Strongly Connected Components}
Consider a directed graph \( G = (V, E) \), where \( V \) is the set of nodes and \( E \subseteq V \times V \) is the set of edges. Two nodes \( u \) and \( w \) in \( G \) are said to be \emph{path equivalent} if there is a path from \( u \) to \( w \) and a path from \( w \) to \( u \). The \emph{strongly connected components} (SCCs) of a graph are the disjoint sets obtained by partitioning \( V \) into sets of path-equivalent vertices\cite{NUUTILA19949}. Since a single node can reach itself (i.e., there is a trivial path from \( u \) to \( u \)), each node represent a \emph{single strongly connected component} (SSCC). In our dataset, the total number of SCCs was expected to be 320,969, the number of nodes in the graph. However, the number of the SCCs in the graph was found to be 320,885 SSCC, along with 62 strongly connected components of size greater than one, which is unreasonable as there must be only one directed edge from a paper to its referenced paper. After analyzing the output, it was found that there are two possible causes for having strongly
connected components with a size greater than one:

\noindent\textbf{Case 1: Academic Journals with Supplement Articles}

Some academic journals publish sets of supplementary articles alongside their regular issues. These supplements often contain special articles, conference proceedings, abstracts, or other materials that align with the main journal’s focus. Although supplements are associated with the primary journal, each article within a supplement is treated as a separate publication and can independently cite any other article, whether within the same supplement or outside it. What distinguishes these articles is the supplement number, which is explicitly stated in the citation.
For example, in the citation “Eur Spine J. 2018 Sep; 27 (Suppl 6)” \cite{globalspine2018}, the term “Suppl 6” indicates that the article is part of Supplement 6 of the journal European Spine Journal. From the output (component number 8,462), we observed a similar case where three papers appear across six supplements of the same journal. These include:
\begin{itemize}
    \item The Global Spine Care Initiative: World Spine Care Executive Summary on Reducing Spine-Related Disability in Low- and Middle-Income Communities. \cite{Haldeman2018ExecutiveSummary}
    \item The Global Spine Care Initiative: Methodology, Contributors, and Disclosures. \cite{Haldeman2018Methodology}
    \item The Global Spine Care Initiative: Model of Care and Implementation. \cite{Johnson2018ModelOfCare}
\end{itemize}
This example highlights how supplementary articles, though part of the same overarching initiative, are published and cited as distinct entities, with their unique supplement identifiers providing clarity.

\noindent\textbf{Case 2: Citation Manipulation Through Pre-Publication Cross-Citation}

There is an adopted behavior by some authors that may seem unethical, where authors cite papers, may be done manually, that are not published yet. Those papers are almost having the same publication year and month as well as some common authors. The target of doing so may be the desire of increasing the number of citations for both papers and consequently increasing the authors' ranks. 

\subsection{Analysis of Co-authorship Network}

\par 
From the analysis, we got insights about the pairs of authors that are frequently cited together (edges in the network) where the weights of the edges  represent the number of times a pair of authors have collaborated or co-authored publications together. These insights include:
\begin{itemize}
    \item Most Frequently Collaborating Pair: The pair of authors A. Hussein (Department of Urology, Cairo University, Egypt) and K. Guru (Roswell Park Cancer Institute) has the highest collaboration frequency, indicated by the maximum weight of 155.
    This suggests that they have collaborated on a significant number of publications, possibly indicating a strong research partnership or shared research interests.
    \item Least Frequently Collaborating Pairs: Within the top 50 pairs, there are seven pairs with a minimum weight of 39. These pairs are A. Soliman (Professor and Consultant Ped Endocrinology Hamad Medical Center, Doha, Qatar) and S. Di Maio (Emeritus Director in Pediatrics, Children's Hospital "Santobono-Pausilipon", Naples, Italy), and H. Elmansy (Urology Department, Northern Ontario School of Medicine, Thunder Bay, Canada) and A. Kotb (Urology Department, Northern Ontario School of Medicine, Thunder Bay, Canada).
    Their relatively lower collaboration frequencies compared to other highly collaborating pairs, indicating fewer instances of joint publications or research collaboration between each pair.
\end{itemize}

\subsubsection{Degree Centrality Measures}

The analysis of the co-authorship network revealed the top 50 authors with the highest degree centrality values, including: D. Nepogodiev (NIHR Academic Clinical Lecturer in Public Health, University of Birmingham), J. Glasbey (NIHR Academic Clinical Lecturer in Global Surgery, NIHR Global Health Research Unit on Global Surgery), A. Bhangu (NIHR Unit on Global Surgery, University of Birmingham, UK), T. Drake (Honorary Clinical Lecturer and Research Fellow, University of Edinburgh), and M. Saad (Guy's and St Thomas's Hospitals, UK). These authors are Egyptian authors, or have extensive collaborations and strong connections with Egyptian authors within the network. Their work is likely to have a significant influence on the field of surgery and public health.
The degree centrality values for these authors are as follows: D. Nepogodiev has a degree centrality of 0.0668, J. Glasbey has a degree centrality of 0.0660, A. Bhangu has a degree centrality of 0.0541, T. Drake has a degree centrality of 0.0538, and M. Saad has a degree centrality of 0.0529.

\subsubsection{Connected Components Analysis}

\par 
The analysis of the coauthorship network revealed a total of 286 connected components, representing groups of authors who have collaborated with each other, forming cohesive subnetworks within the larger coauthorship network.
Each connected component represents a distinct community of authors with strong collaborative ties. The presence of multiple components highlights the fragmented nature of collaborations and the existence of various research communities or subfields within the overall research domain.
\begin{figure}
	\centering 
	\includegraphics[width=0.45\textwidth]{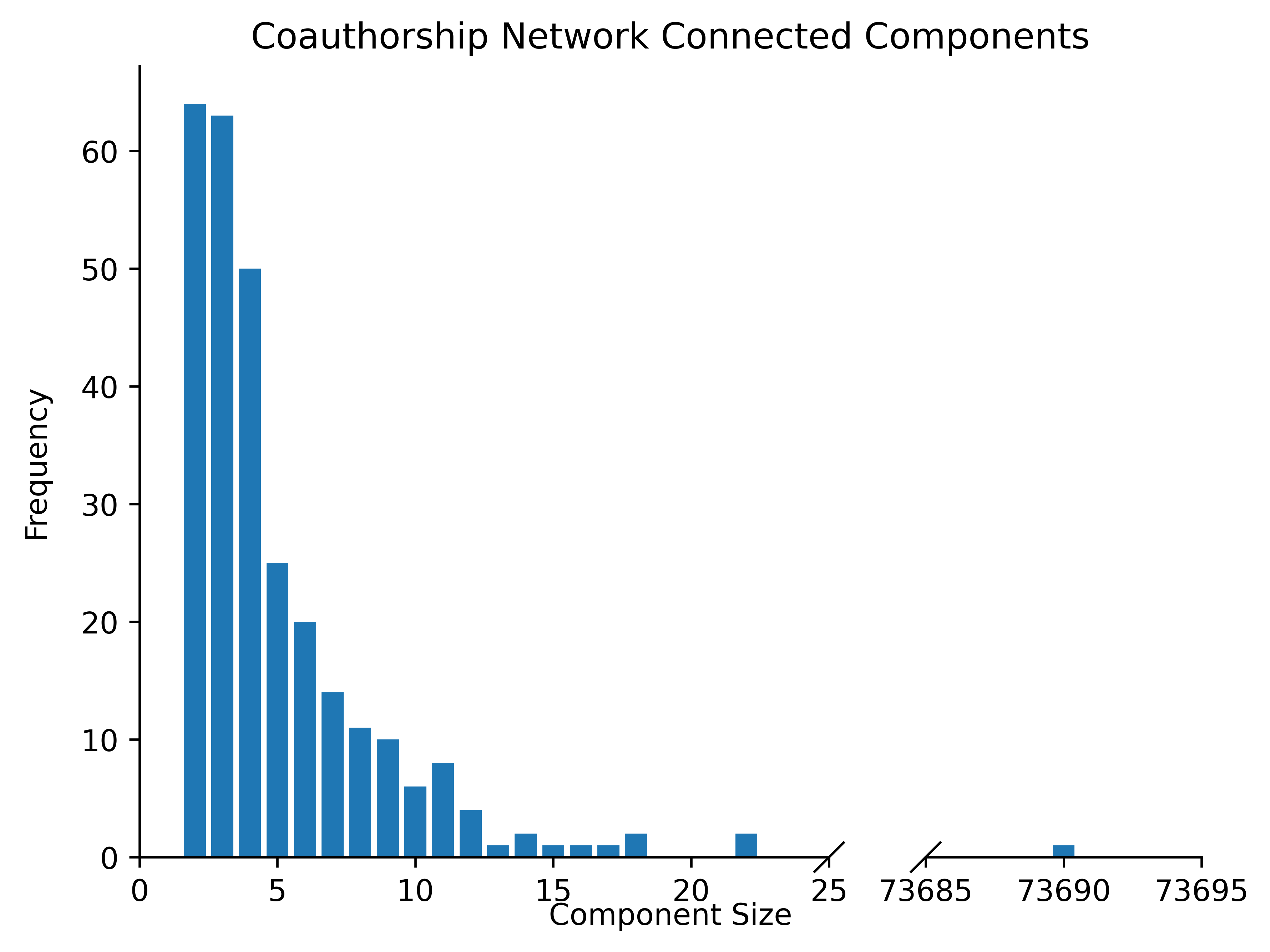}	
	\caption{Connected Components} 
	\label{coauthorconnected}%
\end{figure}
The distribution of the connected components is illustrated in Figure \ref{coauthorconnected}:
\begin{itemize}
    \item The majority of components are small-sized, with sizes ranging from 2 to 5 authors, as shown by the high frequencies on the left-hand side of the distribution. These smaller components reflect localized collaborations, often among close research groups or smaller projects.
    \item A gradual decrease in frequency is observed for larger component sizes, with very few components containing between 10–20 authors.
    \item Notably, there is a single exceptionally large component (73690 authors), likely representing a massive collaboration, such as a consortium, interdisciplinary group, or a globally significant research effort.
\end{itemize}
This trend aligns with the common structure of collaborative networks, where few large communities coexist with many small, isolated groups. By identifying the sizes and distributions of these connected components, we gain valuable insights into the organization of research collaborations.


\begin{figure}
	\centering 
	\includegraphics[width=0.45\textwidth]{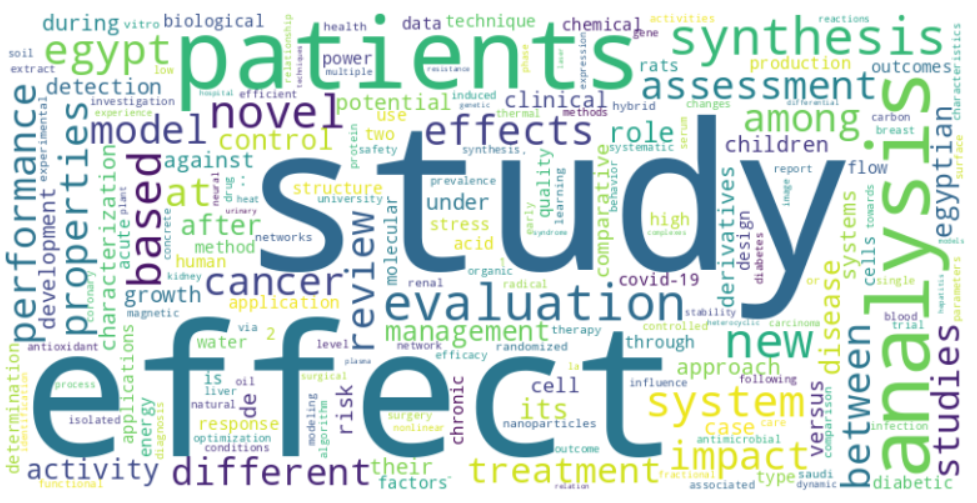}	
	\caption{A word cloud for common words in the papers' titles. } 
	\label{wordcloud}%
\end{figure}

\section{Conclusion}

\par
In this paper, the analysis of the citation network provided insights into the temporal trends of publications, degree distribution, authorship patterns, clustering coefficients, and density. The majority of publications were recently published, indicating a growing research output. The degree distribution followed a power law distribution, with a few highly cited papers and a majority with fewer citations. Collaborative research was observed, with multi-authored papers being the most common. In addition, the analysis of the co-authorship network revealed pairs of authors who frequently collaborate, indicating strong research partnerships. Degree centrality measures identified the top 50 authors with the highest collaboration frequencies, highlighting their influential roles in the research community. Connected components analysis uncovered 286 distinct communities of authors with strong collaborative ties, indicating the presence of various research communities within the broader field. Furthermore, a word cloud of the most common words in the paper titles reveals the central themes and research trends within the analyzed body of work, as shown in figure \ref{wordcloud}. Key terms such as ['study,' 'effect,' 'patients', 'analysis', 'synthesis', 'cancer'] stand out, reflecting the dominant areas of focus and the overall research priorities in different domains. 

\par 
Overall, our analysis provides valuable insights into the publication and collaboration patterns of Egyptian researchers. It highlights the impact of influential publications, the significance of collaborative research, and the presence of diverse research communities within the Egyptian research landscape.
These findings can contribute to a better understanding of the research dynamics in Egypt and facilitate future collaborations and knowledge dissemination within the scientific community.

\bigskip

\section*{Acknowledgements}
We would like to express our deepest gratitude to Eng. Zeyad Shokry for his invaluable guidance and assistance with the collection of our dataset (AlGoNet).

\bigskip

\bibliographystyle {IEEEtran}
\bibliography{ref}

\end{document}